\documentclass[12pt,psfig]{dcds2}

\def\fileversion{v1.20a}
\def\filedate{21.6.94}
\edef\epsfigRestoreAt{\catcode`@=\number\catcode`@\relax}%
\catcode`\@=11\relax
\ifx\undefined\@makeother                
\def\@makeother#1{\catcode`#1=12\relax}  
\fi                                      
\immediate\write16{Document style option `epsfig', \fileversion\space
<\filedate> (edited by SPQR + pks)}
\newcount\EPS@Height \newcount\EPS@Width \newcount\EPS@xscale
\newcount\EPS@yscale
\def\psfigdriver#1{%
  \bgroup\edef\next{\def\noexpand\tempa{#1}}%
    \uppercase\expandafter{\next}%
    \def\LN{DVITOLN03}%
    \def\DVItoPS{DVITOPS}%
    \def\DVIPS{DVIPS}%
    \def\emTeX{EMTEX}%
    \def\OzTeX{OZTEX}%
    \def\Textures{TEXTURES}%
    \global\chardef\fig@driver=0
    \ifx\tempa\LN
        \global\chardef\fig@driver=0\fi
    \ifx\tempa\DVItoPS
        \global\chardef\fig@driver=1\fi
    \ifx\tempa\DVIPS
        \global\chardef\fig@driver=2\fi
    \ifx\tempa\emTeX
        \global\chardef\fig@driver=3\fi
    \ifx\tempa\OzTeX
        \global\chardef\fig@driver=4\fi
    \ifx\tempa\Textures
        \global\chardef\fig@driver=5\fi
  \egroup
\def\psfig@start{}%
\def\psfig@end{}%
\def\epsfig@gofer{}%
\ifcase\fig@driver
\typeout{WARNING! ****
 no specials for LN03 psfig}%
\or 
\def\psfig@start{}%
\def\psfig@end{\special{dvitops: import \@p@sfilefinal \space
\@p@swidth sp \space \@p@sheight sp \space fill}%
\if@clip \typeout{Clipping not supported}\fi
\if@angle \typeout{Rotating not supported}\fi
}%
\let\epsfig@gofer\psfig@end
\or 
\def\psfig@start{\special{ps::[begin]  \@p@swidth \space \@p@sheight \space%
        \@p@sbbllx \space \@p@sbblly \space%
        \@p@sbburx \space \@p@sbbury \space%
        startTexFig \space }%
        \if@clip
                \if@verbose
                        \typeout{(clipped to BB) }%
                \fi
                \special{ps:: doclip \space }%
        \fi
        \if@angle              
                \special {ps:: \@p@sangle \space rotate \space}
        \fi
        \special{ps: plotfile \@p@sfilefinal \space }%
        \special{ps::[end] endTexFig \space }%
}%
\def\psfig@end{}%
\def\epsfig@gofer{\if@clip
                        \if@verbose
                           \typeout{(clipped to BB)}%
                        \fi
                        \epsfclipon
                  \fi
                  \epsfsetgraph{\@p@sfilefinal}%
}%
\or 
\typeout{WARNING. You must have a .bb info file with the Bounding Box
  of the pcx file}%
\def\psfig@start{}%
\def\psfig@end{\typeout{pcx import of \@p@sfilefinal}%
\if@clip \typeout{Clipping not supported}\fi
\if@angle \typeout{Rotating not supported}\fi
\raisebox{\@p@srheight sp}{\special{em: graph \@p@sfilefinal}}}%
\def\epsfig@gofer{}%
\or 
\def\psfig@start{}%
\def\psfig@end{%
\EPS@Width\@p@swidth
\EPS@Height\@p@sheight
\divide\EPS@Width by 65781  
\divide\EPS@Height by 65781
\special{epsf=\@p@sfilefinal
\space
width=\the\EPS@Width
\space
height=\the\EPS@Height
}%
\if@clip \typeout{Clipping not supported}\fi
\if@angle \typeout{Rotating not supported}\fi
}%
\let\epsfig@gofer\psfig@end
\or 
\def\psfig@end{
         \EPS@Width=\@bbw  
         \divide\EPS@Width by 1000
         \EPS@xscale=\@p@swidth \divide \EPS@xscale by \EPS@Width
         \EPS@Height=\@bbh  
         \divide\EPS@Height by 1000
         \EPS@yscale=\@p@sheight \divide \EPS@yscale by\EPS@Height
  \ifnum\EPS@xscale>\EPS@yscale\EPS@xscale=\EPS@yscale\fi
\if@clip
   \if@verbose
      \typeout{(clipped to BB)}%
   \fi
   \epsfclipon
\fi
\special{illustration \@p@sfilefinal\space scaled \the\EPS@xscale}%
}%
\def\psfig@start{}%
\let\epsfig\psfig
\else
\typeout{WARNING. *** unknown  driver - no psfig}%
\fi
}%
\newdimen\ps@dimcent
\ifx\undefined\fbox
\newdimen\fboxrule
\newdimen\fboxsep
\newdimen\ps@tempdima
\newbox\ps@tempboxa
\long\def\fbox#1{\leavevmode\setbox\ps@tempboxa\hbox{#1}\ps@tempdima\fboxrule
    \advance\ps@tempdima \fboxsep \advance\ps@tempdima \dp\ps@tempboxa
   \hbox{\lower \ps@tempdima\hbox
  {\vbox{\hrule height \fboxrule
          \hbox{\vrule width \fboxrule \hskip\fboxsep
          \vbox{\vskip\fboxsep \box\ps@tempboxa\vskip\fboxsep}\hskip
                 \fboxsep\vrule width \fboxrule}%
                 \hrule height \fboxrule}}}}%
\fi
\ifx\@ifundefined\undefined
\long\def\@ifundefined#1#2#3{\expandafter\ifx\csname
  #1\endcsname\relax#2\else#3\fi}%
\fi
\@ifundefined{typeout}%
{\gdef\typeout#1{\immediate\write\sixt@@n{#1}}}%
{\relax}%
\@ifundefined{epsfig}{}{\typeout{EPSFIG --- already loaded}\endinput}%
\@ifundefined{epsfbox}{\input epsf}{}%
\ifx\undefined\@latexerr
        \newlinechar`\^^J
        \def\@spaces{\space\space\space\space}%
        \def\@latexerr#1#2{%
        \edef\@tempc{#2}\expandafter\errhelp\expandafter{\@tempc}%
        \typeout{Error. \space see a manual for explanation.^^J
         \space\@spaces\@spaces\@spaces Type \space H <return> \space for
         immediate help.}\errmessage{#1}}%
\fi
\def\@whattodo{You tried to include a PostScript figure which
cannot be found^^JIf you press return to carry on anyway,^^J
The failed name will be printed in place of the figure.^^J
or type X to quit}%
\def\@whattodobb{You tried to include a PostScript figure which
has no^^Jbounding box, and you supplied none.^^J
If you press return to carry on anyway,^^J
The failed name will be printed in place of the figure.^^J
or type X to quit}%
\def\@nnil{\@nil}%
\def\@empty{}%
\def\@psdonoop#1\@@#2#3{}%
\def\@psdo#1:=#2\do#3{\edef\@psdotmp{#2}\ifx\@psdotmp\@empty \else
    \expandafter\@psdoloop#2,\@nil,\@nil\@@#1{#3}\fi}%
\def\@psdoloop#1,#2,#3\@@#4#5{\def#4{#1}\ifx #4\@nnil \else
       #5\def#4{#2}\ifx #4\@nnil \else#5\@ipsdoloop #3\@@#4{#5}\fi\fi}%
\def\@ipsdoloop#1,#2\@@#3#4{\def#3{#1}\ifx #3\@nnil
       \let\@nextwhile=\@psdonoop \else
      #4\relax\let\@nextwhile=\@ipsdoloop\fi\@nextwhile#2\@@#3{#4}}%
\def\@tpsdo#1:=#2\do#3{\xdef\@psdotmp{#2}\ifx\@psdotmp\@empty \else
    \@tpsdoloop#2\@nil\@nil\@@#1{#3}\fi}%
\def\@tpsdoloop#1#2\@@#3#4{\def#3{#1}\ifx #3\@nnil
       \let\@nextwhile=\@psdonoop \else
      #4\relax\let\@nextwhile=\@tpsdoloop\fi\@nextwhile#2\@@#3{#4}}%
\long\def\epsfaux#1#2:#3\\{\ifx#1\epsfpercent
   \def\testit{#2}\ifx\testit\epsfbblit
        \@atendfalse
        \epsf@atend #3 . \\%
        \if@atend
           \if@verbose
                \typeout{epsfig: found `(atend)'; continuing search}%
           \fi
        \else
                \epsfgrab #3 . . . \\%
                \epsffileokfalse\global\no@bbfalse
                \global\epsfbbfoundtrue
        \fi
   \fi\fi}%
\def\epsf@atendlit{(atend)}
\def\epsf@atend #1 #2 #3\\{%
   \def\epsf@tmp{#1}\ifx\epsf@tmp\empty
      \epsf@atend #2 #3 .\\\else
   \ifx\epsf@tmp\epsf@atendlit\@atendtrue\fi\fi}%

\chardef\trig@letter = 11
\chardef\other = 12
 
\newif\ifdebug 
\newif\ifc@mpute 
\newif\if@atend
\c@mputetrue 
 
\let\then = \relax
\def\r@dian{pt }%
\let\r@dians = \r@dian
\let\dimensionless@nit = \r@dian
\let\dimensionless@nits = \dimensionless@nit
\def\internal@nit{sp }%
\let\internal@nits = \internal@nit
\newif\ifstillc@nverging
\def \Mess@ge #1{\ifdebug \then \message {#1} \fi}%
 
{ 
        \catcode `\@ = \trig@letter
        \gdef \nodimen {\expandafter \n@dimen \the \dimen}%
        \gdef \term #1 #2 #3%
               {\edef \t@ {\the #1}
                \edef \t@@ {\expandafter \n@dimen \the #2\r@dian}%
                \t@rm {\t@} {\t@@} {#3}%
               }%
        \gdef \t@rm #1 #2 #3%
               {{%
                \count 0 = 0
                \dimen 0 = 1 \dimensionless@nit
                \dimen 2 = #2\relax
                \Mess@ge {Calculating term #1 of \nodimen 2}%
                \loop
                \ifnum  \count 0 < #1
                \then   \advance \count 0 by 1
                        \Mess@ge {Iteration \the \count 0 \space}%
                        \Multiply \dimen 0 by {\dimen 2}%
                        \Mess@ge {After multiplication, term = \nodimen 0}%
                        \Divide \dimen 0 by {\count 0}%
                        \Mess@ge {After division, term = \nodimen 0}%
                \repeat
                \Mess@ge {Final value for term #1 of
                                \nodimen 2 \space is \nodimen 0}%
                \xdef \Term {#3 = \nodimen 0 \r@dians}%
                \aftergroup \Term
               }}%
        \catcode `\p = \other
        \catcode `\t = \other
        \gdef \n@dimen #1pt{#1} 
}%
 
\def \Divide #1by #2{\divide #1 by #2} 
 
\def \Multiply #1by #2
       {{
        \count 0 = #1\relax
        \count 2 = #2\relax
        \count 4 = 65536
        \Mess@ge {Before scaling, count 0 = \the \count 0 \space and
                        count 2 = \the \count 2}%
        \ifnum  \count 0 > 32767 
        \then   \divide \count 0 by 4
                \divide \count 4 by 4
        \else   \ifnum  \count 0 < -32767
                \then   \divide \count 0 by 4
                        \divide \count 4 by 4
                \else
                \fi
        \fi
        \ifnum  \count 2 > 32767 
        \then   \divide \count 2 by 4
                \divide \count 4 by 4
        \else   \ifnum  \count 2 < -32767
                \then   \divide \count 2 by 4
                        \divide \count 4 by 4
                \else
                \fi
        \fi
        \multiply \count 0 by \count 2
        \divide \count 0 by \count 4
        \xdef \product {#1 = \the \count 0 \internal@nits}%
        \aftergroup \product
       }}%
 
\def\r@duce{\ifdim\dimen0 > 90\r@dian \then   
                \multiply\dimen0 by -1
                \advance\dimen0 by 180\r@dian
                \r@duce
            \else \ifdim\dimen0 < -90\r@dian \then  
                \advance\dimen0 by 360\r@dian
                \r@duce
                \fi
            \fi}%
 
\def\Sine#1%
       {{%
        \dimen 0 = #1 \r@dian
        \r@duce
        \ifdim\dimen0 = -90\r@dian \then
           \dimen4 = -1\r@dian
           \c@mputefalse
        \fi
        \ifdim\dimen0 = 90\r@dian \then
           \dimen4 = 1\r@dian
           \c@mputefalse
        \fi
        \ifdim\dimen0 = 0\r@dian \then
           \dimen4 = 0\r@dian
           \c@mputefalse
        \fi
        \ifc@mpute \then
                \divide\dimen0 by 180
                \dimen0=3.141592654\dimen0
                \dimen 2 = 3.1415926535897963\r@dian 
                \divide\dimen 2 by 2 
                \Mess@ge {Sin: calculating Sin of \nodimen 0}%
                \count 0 = 1 
                \dimen 2 = 1 \r@dian 
                \dimen 4 = 0 \r@dian 
                \loop
                        \ifnum  \dimen 2 = 0 
                        \then   \stillc@nvergingfalse
                        \else   \stillc@nvergingtrue
                        \fi
                        \ifstillc@nverging 
                        \then   \term {\count 0} {\dimen 0} {\dimen 2}%
                                \advance \count 0 by 2
                                \count 2 = \count 0
                                \divide \count 2 by 2
                                \ifodd  \count 2 
                                \then   \advance \dimen 4 by \dimen 2
                                \else   \advance \dimen 4 by -\dimen 2
                                \fi
                \repeat
        \fi
                        \xdef \sine {\nodimen 4}%
       }}%
 
\def\Cosine#1{\ifx\sine\UnDefined\edef\Savesine{\relax}\else
                             \edef\Savesine{\sine}\fi
        {\dimen0=#1\r@dian\multiply\dimen0 by -1
         \advance\dimen0 by 90\r@dian
         \Sine{\nodimen 0}%
         \xdef\cosine{\sine}%
         \xdef\sine{\Savesine}}}
\def\psdraft{\def\@psdraft{0}}%
\def\psfull{\def\@psdraft{1}}%
\psfull
\newif\if@compress
\def\pscompress{\@compresstrue}
\def\psnocompress{\@compressfalse}
\@compressfalse
\newif\if@scalefirst
\def\psscalefirst{\@scalefirsttrue}%
\def\psrotatefirst{\@scalefirstfalse}%
\psrotatefirst
\newif\if@draftbox
\def\psnodraftbox{\@draftboxfalse}%
\@draftboxtrue
\newif\if@noisy
\@noisyfalse
\newif\ifno@bb
\newif\if@bbllx
\newif\if@bblly
\newif\if@bburx
\newif\if@bbury
\newif\if@height
\newif\if@width
\newif\if@rheight
\newif\if@rwidth
\newif\if@angle
\newif\if@clip
\newif\if@verbose
\newif\if@prologfile
\def\@p@@sprolog#1{\@prologfiletrue\def\@prologfileval{#1}}%
\def\@p@@sclip#1{\@cliptrue}%
\newif\ifepsfig@dos  
\def\epsfigdos{\epsfig@dostrue}%
\epsfig@dosfalse
\newif\ifuse@psfig
\def\ParseName#1{\expandafter\@Parse#1}%
\def\@Parse#1.#2:{\gdef\BaseName{#1}\gdef\FileType{#2}}%

\def\@p@@sfile#1{%
  \ifepsfig@dos
     \ParseName{#1:}%
  \else
     \gdef\BaseName{#1}\gdef\FileType{}%
  \fi
  \def\@p@sfile{NO FILE: #1}%
  \def\@p@sfilefinal{NO FILE: #1}%
  \openin1=#1
  \ifeof1\closein1\openin1=\BaseName.bb
    \ifeof1\closein1
      \if@bbllx                 
        \if@bblly\if@bburx\if@bbury
          \def\@p@sfile{#1}%
          \def\@p@sfilefinal{#1}%
        \fi\fi\fi
      \else                     
        \@latexerr{ERROR. PostScript file #1 not found}\@whattodo
        \@p@@sbbllx{100bp}%
        \@p@@sbblly{100bp}%
        \@p@@sbburx{200bp}%
        \@p@@sbbury{200bp}%
        \psdraft
      \fi
    \else                       
      \closein1%
      \edef\@p@sfile{\BaseName.bb}%
      \typeout{using BB from \@p@sfile}%
      \ifnum\fig@driver=3
        \edef\@p@sfilefinal{\BaseName.pcx}%
      \else
        \ifepsfig@dos
          \edef\@p@sfilefinal{"`gunzip -c `texfind \BaseName.{z,Z,gz}"}%
        \else
          \edef\@p@sfilefinal{"`epsfig \if@compress-c \fi#1"}%
        \fi
      \fi
    \fi
  \else\closein1                
    \edef\@p@sfile{#1}%
    \if@compress  
      \edef\@p@sfilefinal{"`epsfig -c #1"}%
    \else
      \edef\@p@sfilefinal{#1}%
    \fi
  \fi%
}

\let\@p@@sfigure\@p@@sfile
\def\@p@@sbbllx#1{%
                                            \@bbllxtrue
                \ps@dimcent=#1
                \edef\@p@sbbllx{\number\ps@dimcent}%
                \divide\ps@dimcent by65536
                \global\edef\epsfllx{\number\ps@dimcent}%
}%
\def\@p@@sbblly#1{%
                \@bbllytrue
                \ps@dimcent=#1
                \edef\@p@sbblly{\number\ps@dimcent}%
                \divide\ps@dimcent by65536
                \global\edef\epsflly{\number\ps@dimcent}%
}%
\def\@p@@sbburx#1{%
                \@bburxtrue
                \ps@dimcent=#1
                \edef\@p@sbburx{\number\ps@dimcent}%
                \divide\ps@dimcent by65536
                \global\edef\epsfurx{\number\ps@dimcent}%
}%
\def\@p@@sbbury#1{%
                \@bburytrue
                \ps@dimcent=#1
                \edef\@p@sbbury{\number\ps@dimcent}%
                \divide\ps@dimcent by65536
                \global\edef\epsfury{\number\ps@dimcent}%
}%
\def\@p@@sheight#1{%
                \@heighttrue
                \global\epsfysize=#1
                \ps@dimcent=#1
                \edef\@p@sheight{\number\ps@dimcent}%
}%
\def\@p@@swidth#1{%
                \@widthtrue
                \global\epsfxsize=#1
                \ps@dimcent=#1
                \edef\@p@swidth{\number\ps@dimcent}%
}%
\def\@p@@srheight#1{%
                \@rheighttrue\use@psfigtrue
                \ps@dimcent=#1
                \edef\@p@srheight{\number\ps@dimcent}%
}%
\def\@p@@srwidth#1{%
                \@rwidthtrue\use@psfigtrue
                \ps@dimcent=#1
                \edef\@p@srwidth{\number\ps@dimcent}%
}%
\def\@p@@sangle#1{%
                \use@psfigtrue
                \@angletrue
                \edef\@p@sangle{#1}%
}%
\def\@p@@ssilent#1{%
                \@verbosefalse
}%
\def\@p@@snoisy#1{%
                \@verbosetrue
}%
\def\@cs@name#1{\csname #1\endcsname}%
\def\@setparms#1=#2,{\@cs@name{@p@@s#1}{#2}}%
\def\ps@init@parms{%
                \@bbllxfalse \@bbllyfalse
                \@bburxfalse \@bburyfalse
                \@heightfalse \@widthfalse
                \@rheightfalse \@rwidthfalse
                \def\@p@sbbllx{}\def\@p@sbblly{}%
                \def\@p@sbburx{}\def\@p@sbbury{}%
                \def\@p@sheight{}\def\@p@swidth{}%
                \def\@p@srheight{}\def\@p@srwidth{}%
                \def\@p@sangle{0}%
                \def\@p@sfile{}%
                \use@psfigfalse
                \@prologfilefalse
                \def\@sc{}%
                \if@noisy
                        \@verbosetrue
                \else
                        \@verbosefalse
                \fi
                \@clipfalse
}%
\def\parse@ps@parms#1{%
                \@psdo\@psfiga:=#1\do
                   {\expandafter\@setparms\@psfiga,}%
\if@prologfile
\fi
}%
\def\bb@missing{%
        \if@verbose
            \typeout{psfig: searching \@p@sfile \space  for bounding box}%
        \fi
        \epsfgetbb{\@p@sfile}%
        \ifepsfbbfound
            \ps@dimcent=\epsfllx bp\edef\@p@sbbllx{\number\ps@dimcent}%
            \ps@dimcent=\epsflly bp\edef\@p@sbblly{\number\ps@dimcent}%
            \ps@dimcent=\epsfurx bp\edef\@p@sbburx{\number\ps@dimcent}%
            \ps@dimcent=\epsfury bp\edef\@p@sbbury{\number\ps@dimcent}%
        \else
            \epsfbbfoundfalse
        \fi
}
\newdimen\p@intvaluex
\newdimen\p@intvaluey
\def\rotate@#1#2{{\dimen0=#1 sp\dimen1=#2 sp
                  \global\p@intvaluex=\cosine\dimen0
                  \dimen3=\sine\dimen1
                  \global\advance\p@intvaluex by -\dimen3
                  \global\p@intvaluey=\sine\dimen0
                  \dimen3=\cosine\dimen1
                  \global\advance\p@intvaluey by \dimen3
                  }}%
\def\compute@bb{%
                \epsfbbfoundfalse
                \if@bbllx\epsfbbfoundtrue\fi
                \if@bblly\epsfbbfoundtrue\fi
                \if@bburx\epsfbbfoundtrue\fi
                \if@bbury\epsfbbfoundtrue\fi
                \ifepsfbbfound\else\bb@missing\fi
                \ifepsfbbfound\else
                \@latexerr{ERROR. cannot locate BoundingBox}\@whattodobb
                        \@p@@sbbllx{100bp}%
                        \@p@@sbblly{100bp}%
                        \@p@@sbburx{200bp}%
                        \@p@@sbbury{200bp}%
                        \no@bbtrue
                        \psdraft
                \fi
                \count203=\@p@sbburx
                \count204=\@p@sbbury
                \advance\count203 by -\@p@sbbllx
                \advance\count204 by -\@p@sbblly
                \edef\ps@bbw{\number\count203}%
                \edef\ps@bbh{\number\count204}%
                 \edef\@bbw{\number\count203}%
                \edef\@bbh{\number\count204}%
               \if@angle
                        \Sine{\@p@sangle}\Cosine{\@p@sangle}%
 
{\ps@dimcent=\maxdimen\xdef\r@p@sbbllx{\number\ps@dimcent}%
 
\xdef\r@p@sbblly{\number\ps@dimcent}%
 
\xdef\r@p@sbburx{-\number\ps@dimcent}%
 
\xdef\r@p@sbbury{-\number\ps@dimcent}}%
                        \def\minmaxtest{%
                           \ifnum\number\p@intvaluex<\r@p@sbbllx
                              \xdef\r@p@sbbllx{\number\p@intvaluex}\fi
                           \ifnum\number\p@intvaluex>\r@p@sbburx
                              \xdef\r@p@sbburx{\number\p@intvaluex}\fi
                           \ifnum\number\p@intvaluey<\r@p@sbblly
                              \xdef\r@p@sbblly{\number\p@intvaluey}\fi
                           \ifnum\number\p@intvaluey>\r@p@sbbury
                              \xdef\r@p@sbbury{\number\p@intvaluey}\fi
                           }%
                        \rotate@{\@p@sbbllx}{\@p@sbblly}%
                        \minmaxtest
                        \rotate@{\@p@sbbllx}{\@p@sbbury}%
                        \minmaxtest
                        \rotate@{\@p@sbburx}{\@p@sbblly}%
                        \minmaxtest
                        \rotate@{\@p@sbburx}{\@p@sbbury}%
                        \minmaxtest
 
\edef\@p@sbbllx{\r@p@sbbllx}\edef\@p@sbblly{\r@p@sbblly}%
 
\edef\@p@sbburx{\r@p@sbburx}\edef\@p@sbbury{\r@p@sbbury}%
                \fi
                \count203=\@p@sbburx
                \count204=\@p@sbbury
                \advance\count203 by -\@p@sbbllx
                \advance\count204 by -\@p@sbblly
                \edef\@bbw{\number\count203}%
                \edef\@bbh{\number\count204}%
}%
\def\in@hundreds#1#2#3{\count240=#2 \count241=#3
                     \count100=\count240        
                     \divide\count100 by \count241
                     \count101=\count100
                     \multiply\count101 by \count241
                     \advance\count240 by -\count101
                     \multiply\count240 by 10
                     \count101=\count240        
                     \divide\count101 by \count241
                     \count102=\count101
                     \multiply\count102 by \count241
                     \advance\count240 by -\count102
                     \multiply\count240 by 10
                     \count102=\count240        
                     \divide\count102 by \count241
                     \count200=#1\count205=0
                     \count201=\count200
                        \multiply\count201 by \count100
                        \advance\count205 by \count201
                     \count201=\count200
                        \divide\count201 by 10
                        \multiply\count201 by \count101
                        \advance\count205 by \count201
                     \count201=\count200
                        \divide\count201 by 100
                        \multiply\count201 by \count102
                        \advance\count205 by \count201
                     \edef\@result{\number\count205}%
}%
\def\compute@wfromh{%
                \in@hundreds{\@p@sheight}{\@bbw}{\@bbh}%
                \edef\@p@swidth{\@result}%
}%
\def\compute@hfromw{%
                \in@hundreds{\@p@swidth}{\@bbh}{\@bbw}%
                \edef\@p@sheight{\@result}%
}%
\def\compute@handw{%
                \if@height
                        \if@width
                        \else
                                \compute@wfromh
                        \fi
                \else
                        \if@width
                                \compute@hfromw
                        \else
                                \edef\@p@sheight{\@bbh}%
                                \edef\@p@swidth{\@bbw}%
                        \fi
                \fi
}%
\def\compute@resv{%
                \if@rheight \else \edef\@p@srheight{\@p@sheight} \fi
                \if@rwidth \else \edef\@p@srwidth{\@p@swidth} \fi
}%
\def\compute@sizes{%
        \if@scalefirst\if@angle
        \if@width
           \in@hundreds{\@p@swidth}{\@bbw}{\ps@bbw}%
           \edef\@p@swidth{\@result}%
        \fi
        \if@height
           \in@hundreds{\@p@sheight}{\@bbh}{\ps@bbh}%
           \edef\@p@sheight{\@result}%
        \fi
        \fi\fi
        \compute@handw
        \compute@resv
}
\long\def\graphic@verb#1{\def\next{#1}%
  {\expandafter\graphic@strip\meaning\next}}
\def\graphic@strip#1>{}
\def\graphic@zapspace#1{%
  #1\ifx\graphic@zapspace#1\graphic@zapspace%
  \else\expandafter\graphic@zapspace%
  \fi}
\def\psfig#1{%
\edef\@tempa{\graphic@zapspace#1{}}%
\ifvmode\leavevmode\fi\vbox {%
        \ps@init@parms
        \parse@ps@parms{\@tempa}%
        \ifnum\@psdraft=1
                \typeout{[\@p@sfilefinal]}%
                \if@verbose
                        \typeout{epsfig: using PSFIG macros}%
                \fi
                \psfig@method
        \else
                \epsfig@draft
        \fi
}
}%
\def\graphic@zapspace#1{%
  #1\ifx\graphic@zapspace#1\graphic@zapspace%
  \else\expandafter\graphic@zapspace%
  \fi}
\def\epsfig#1{%
\edef\@tempa{\graphic@zapspace#1{}}%
\ifvmode\leavevmode\fi\vbox {%
        \ps@init@parms
        \parse@ps@parms{\@tempa}%
        \ifnum\@psdraft=1
          \if@angle\use@psfigtrue\fi
          {\ifnum\fig@driver=1\global\use@psfigtrue\fi}%
          {\ifnum\fig@driver=3\global\use@psfigtrue\fi}%
          {\ifnum\fig@driver=4\global\use@psfigtrue\fi}%
          {\ifnum\fig@driver=5\global\use@psfigtrue\fi}%
                \ifuse@psfig
                        \if@verbose
                                \typeout{epsfig: using PSFIG macros}%
                        \fi
                        \psfig@method
                \else
                        \if@verbose
                                \typeout{epsfig: using EPSF macros}%
                        \fi
                        \epsf@method
                \fi
        \else
                \epsfig@draft
        \fi
}%
}%

\def\epsf@method{%
        \epsfbbfoundfalse
        \if@bbllx\epsfbbfoundtrue\fi
        \if@bblly\epsfbbfoundtrue\fi
        \if@bburx\epsfbbfoundtrue\fi
        \if@bbury\epsfbbfoundtrue\fi
        \ifepsfbbfound\else\epsfgetbb{\@p@sfile}\fi
        \ifepsfbbfound
           \typeout{<\@p@sfilefinal>}%
           \epsfig@gofer
        \else
          \@latexerr{ERROR - Cannot locate BoundingBox}\@whattodobb
          \@p@@sbbllx{100bp}%
          \@p@@sbblly{100bp}%
          \@p@@sbburx{200bp}%
          \@p@@sbbury{200bp}%
                \count203=\@p@sbburx
                \count204=\@p@sbbury
                \advance\count203 by -\@p@sbbllx
                \advance\count204 by -\@p@sbblly
                \edef\@bbw{\number\count203}%
                \edef\@bbh{\number\count204}%
          \compute@sizes
          \epsfig@@draft
       \fi
}%
\def\psfig@method{%
        \compute@bb
        \ifepsfbbfound
          \compute@sizes
          \psfig@start
          \vbox to \@p@srheight sp{\hbox to \@p@srwidth 
            sp{\hss}\vss\psfig@end}%
        \else
           \epsfig@draft
        \fi
}%
\def\epsfig@draft{\compute@bb\compute@sizes\epsfig@@draft}%
\def\epsfig@@draft{%
\typeout{<(draft only) \@p@sfilefinal>}%
\if@draftbox
        \hbox{{\fboxsep0pt\fbox{\vbox to \@p@srheight sp{%
        \vss\hbox to \@p@srwidth sp{ \hss 
           \expandafter\Literally\@p@sfilefinal\@nil
                          \hss }\vss
        }}}}%
\else
        \vbox to \@p@srheight sp{%
        \vss\hbox to \@p@srwidth sp{\hss}\vss}%
\fi
}%
\def\Literally#1\@nil{{\tt\graphic@verb{#1}}}
\psfigdriver{dvips}%
\epsfigRestoreAt

\def\ppages{1--10}

\theoremstyle{plain}

\theoremstyle{definition}

\title[Classical and Quantum Chaos in Fundamental Field Theories]
 {Classical and Quantum Chaos in Fundamental Field Theories}


\author[Harald Markum and Rainer Pullirsch]
       {}

\begin{document}
\maketitle

\centerline{\scshape Harald Markum and Rainer Pullirsch}
\medskip

{\footnotesize \centerline{} \centerline{
Atominstitut der \"osterreichischen Universit\"aten}
\centerline{ Technische Universit\"at Wien}
\centerline{ Wiedner Hauptstra\ss e 8-10}
\centerline{ A-1040 Vienna, Austria}
 }

\begin{quote}{\normalfont\fontsize{8}{10}\selectfont
{\bfseries Abstract.}
An investigation of classical chaos and quantum chaos in 
gauge fields and fermion fields, respectively, is presented
for (quantum) electrodynamics.
We analyze the leading Lyapunov exponents of U(1)
gauge field configurations on a $12^3$ lattice
which are initialized by Monte Carlo simulations.
We find that configurations in the strong coupling phase are
substantially more chaotic than in the deconfinement phase.
Considering the quantum case, 
complete eigenvalue spectra of the Dirac operator in
quenched $4d$ compact QED are studied on $8^3 \times 4$ and $8^3 \times 6$
lattices. We investigate the behavior of the nearest-neighbor
spacing distribution $P(s)$ as a measure of the fluctuation
properties of the eigenvalues in the strong coupling and the
Coulomb phase. In both phases we find agreement with the Wigner
surmise of the unitary ensemble of random-matrix theory
indicating quantum chaos.
\par}
\end{quote}

\section{Lyapunov exponents in Minkowskian U(1) gauge theory}

\subsection{Classical chaotic dynamics from Monte Carlo initial states}

Cha\-otic dynamics in general is characterized by the
spectrum of Lyapunov exponents. These exponents, if they are positive,
reflect an exponential divergence of initially adjacent configurations.
In case of symmetries inherent in the Hamiltonian of the system
there are corresponding zero values of these exponents. Finally
negative exponents belong to irrelevant directions in the phase
space: perturbation components in these directions die out
exponentially. Pure gauge fields on the lattice show a characteristic
Lyapunov spectrum consisting of one third of each kind of
exponents~\cite{BOOK}.
Assuming this general structure of the Lyapunov spectrum we
investigate presently its magnitude only, namely the maximal
value of the Lyapunov exponent, $L_{{\rm max}}$.

The general definition of the Lyapunov exponent is based on a
distance measure $d(t)$ in phase space,
\begin{equation}
L := \lim_{t\rightarrow\infty} \lim_{d(0)\rightarrow 0}
\frac{1}{t} \ln \frac{d(t)}{d(0)}.
\end{equation}
In case of conservative dynamics the sum of all Lyapunov exponents
is zero according to Liouville's theorem, $\sum L_i = 0$.
We utilize the gauge invariant distance measure consisting of
the local differences of energy densities between two $3d$ field configurations
on the lattice:
\begin{equation}
d : = \frac{1}{N_P} \sum_P\nolimits \, \left| {\rm tr} U_P - {\rm tr} U'_P \right|.
\end{equation}
Here the symbol $\sum_P$ stands for the sum over all $N_P$ plaquettes,
so this distance is bound in the interval $(0,2N)$ for the group
SU(N). $U_P$ and $U'_P$ are the plaquette variables, constructed from
the basic link variables $U_{x,i}$,
\begin{equation}
U_{x,i} = \exp \left( aA_{x,i}^cT^c \right)\: ,
\end{equation}
located on lattice links pointing from the position $x=(x_1,x_2,x_3)$ to
$x+ae_i$. The generators of the group are
$T^c = -ig\tau^c/2$ with $\tau^c$ being the Pauli matrices
in case of SU(2) and $A_{x,i}^c$ is the vector potential.
The elementary plaquette variable is constructed for a plaquette with a
corner at $x$ and lying in the $ij$-plane as
$U_{x,ij} = U_{x,i} U_{x+i,j} U^{\dag}_{x+j,i} U^{\dag}_{x,j}$.
It is related to the magnetic field strength $B_{x,k}^c$:
\begin{equation}
U_{x,ij} = \exp \left( \varepsilon_{ijk} a B_{x,k}^c T^c \right).
\end{equation}
The electric field strength $E_{x,i}^c$ is related to the canonically conjugate
momentum $P_{x,i} = \dot{U}_{x,i}$ via
\begin{equation}
E^c_{x,i} = \frac{2a}{g^3} {\rm tr} \left( T^c \dot{U}_{x,i} U^{\dag}_{x,i} \right).
\end{equation}

The Hamiltonian of the lattice gauge field system can be casted into
the form
\begin{equation}
H = \sum \left[ \frac{1}{2} \langle P, P \rangle \, + \,
 1 - \frac{1}{4} \langle U, V \rangle \right].
\end{equation}
Here the scalar product stands for
$\langle A, B \rangle = \frac{1}{2} {\rm tr} (A B^{\dag} )$.
The staple variable $V$ is a sum of triple products of elementary
link variables closing a plaquette with the chosen link $U$.
This way the Hamiltonian is formally written as a sum over link
contributions and $V$ plays the role of the classical force
acting on the link variable $U$. 

Initial conditions chosen randomly with a given average magnetic energy
per plaquette have been investigated in past years~\cite{Biro94}.
We prepare the initial field configurations
from a standard four dimensional Euclidean Monte Carlo program on
a $12^3\times 4$ lattice varying the gauge coupling $g$~\cite{SU2}.
We relate such four dimensional Euclidean
lattice field configurations to Minkow\-skian momenta and fields
for the three dimensional Hamiltonian simulation
by identifying a fixed time slice of the four dimensional lattice.

\begin{figure}[t]
\centerline{{\psfig{figure=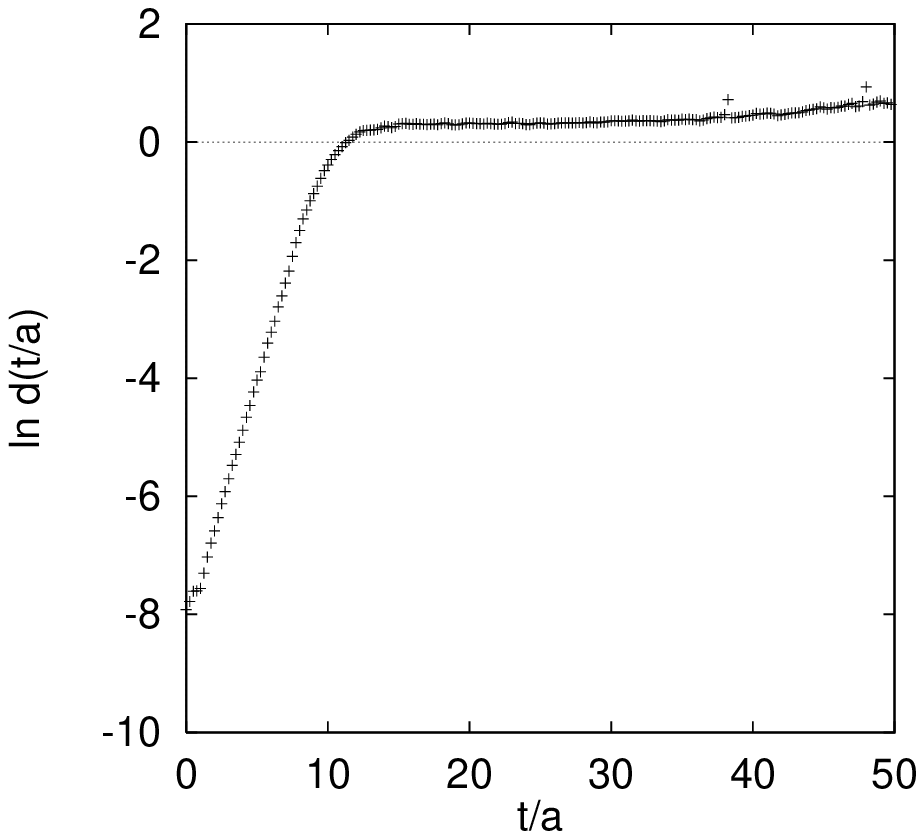,width=6cm}}\hspace{5mm}
{\psfig{figure=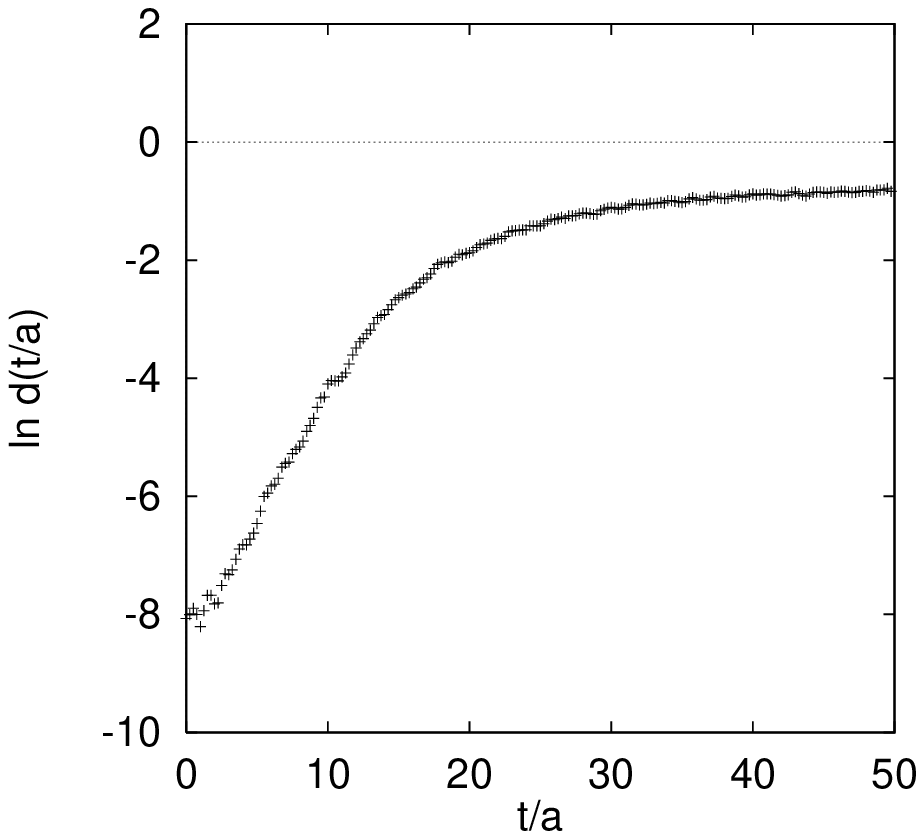,width=6cm}}}
\caption{
  Exponentially diverging distance in real time of initially adjacent U(1) field
  configurations on a $12^3$ lattice prepared at $\beta=0.9$ in the
  confinement (left) and at $\beta=1.1$ in the Coulomb
  phase (right).
\label{fig1}
 }
\end{figure}

\subsection{Chaos, confinement and continuum limit}

We start the presentation of our results with a characteristic example
of the time evolution of the distance between initially adjacent
configurations. An initial state prepared by a standard four dimensional
Monte Carlo simulation is evolved according to the classical Hamiltonian dynamics
in real time. Afterwards this initial state is rotated locally by
group elements which are chosen randomly near to the unity.
The time evolution of this slightly rotated configuration is then
pursued and finally the distance between these two evolutions
is calculated at the corresponding times.
A typical exponential rise of this distance followed by a saturation
can be inspected in Fig.~\ref{fig1} from an example of U(1) gauge theory
for two values of $\beta=1/g^2$ in the confinement phase and in the Coulomb phase.
While the saturation is an artifact of
the compact distance measure of the lattice, the exponential rise
(the linear rise of the logarithm)
can be used for the determination of the leading Lyapunov exponent.
The left plot exhibits that in the confinement phase the
field has larger Lyapunov exponents than in the Coulomb phase 
shown in the right plot.

\begin{figure}[t]
\centerline{\psfig{figure=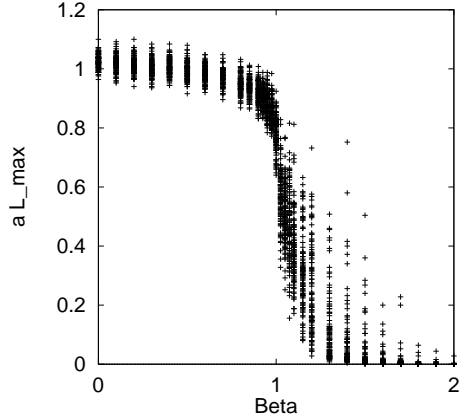,width=6cm}}
\caption{Lyapunov exponents of 100 U(1) field configurations as a function
         of coupling.
\label{fig2}
 }
\end{figure}

The main result of the present study is the dependence of the leading
Lyapunov exponent $L_{{\rm max}}$ on the inverse coupling strength $\beta$,
displayed in Fig.~\ref{fig2}.
As expected the strong coupling phase
is more chaotic.  The transition reflects
the critical coupling to the Coulomb phase.

\begin{figure}[t]
\centerline{\psfig{figure=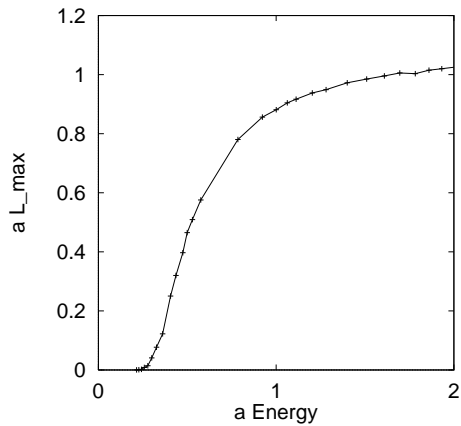,width=6cm}}
\caption{
  Average maximal Lyapunov exponent as a function of the
  scaled average energy per plaquette $ag^2E$. The U(1) gauge theory
  shows an approximately quadratic behavior in the weak coupling regime.
\label{fig3}
 }
\end{figure}

An interesting result concerning the continuum limit can be viewed from Fig.~\ref{fig3}
which shows the energy dependence of the Lyapunov exponents for the U(1) theory.
One observes an approximately quadratic relation in the weak coupling regime.
From scaling arguments one expects a functional relationship between
the Lyapunov exponent and the energy \cite{BOOK,SCALING}
\begin{equation}
L(a) \propto a^{k-1} E^{k}(a) ,
\label{scaling}
\end{equation}
with the exponent $k$ being crucial for the continuum limit of the
classical field theory. A value of $k < 1$ leads to a
divergent Lyapunov exponent, while $k > 1$ yields a vanishing $L$ in
the continuum. The case $k = 1$ is special allowing for a finite non-zero
Lyapunov exponent. Our analysis of the scaling relation (\ref{scaling})
gives evidence that the classical compact U(1) lattice gauge theory
has $k \approx 2$ and with $L(a) \to 0$ a regular continuum theory. 

\section{Quantum chaos in compact Euclidean QED}

\subsection{Manifestation of quantum chaos} 

The fluctuation properties of the eigenvalues of Dirac operator
for quantum chromodynamics (QCD) on a lattice in Euclidean space-time
have attracted much attention in the past few years. In 
Ref.~\cite{Verb95} it was first shown for SU(2) lattice gauge theory 
that certain features of the spectrum of the Dirac operator are described
by random-matrix theory (RMT). In particular the
so-called nearest-neighbor spacing distribution $P(s)$, i.e. the distribution
of the spacings $s$ of adjacent eigenvalues on the ``unfolded'' scale,
agrees with the Wigner surmise of RMT. According to the 
Bohigas-Giannoni-Schmit conjecture \cite{Bohi84}, quantum systems whose
classical counterparts are chaotic have a $P(s)$ given by RMT whereas systems
whose classical counterparts are integrable obey a Poisson distribution
$P(s)=e^{-s}$. Therefore, the specific form of $P(s)$ is often taken as a
criterion for ``quantum chaos''. However, there is no accepted proof of the
Bohigas-Giannoni-Schmit conjecture yet. The field of quantum chaos is
still developing and there are many open conceptual problems \cite{riew}.
Applying this conjecture it was recently demonstrated that QCD is chaotic, both
in the confinement and the quark gluon plasma phase \cite{Pull98}.
 
A number of interesting results have been established for
chaotic dynamics in classical gauge theories. Lattice gauge
theories are chaotic as classical Hamiltonian dynamical systems 
\cite{Biro94}. Furthermore, it was found that the leading Lyapunov exponent
of SU(2) Yang-Mills field configurations indicates that configurations
corresponding to
the deconfinement phase are chaotic although they are less chaotic
than in the strong coupling phase at finite temperature \cite{SU2}.
The scaling of the maximal Lyapunov exponent in the classical 
continuum limit was studied in Ref.~\cite{SCALING}: It was
suggested that Abelian gauge theories behave regularly in the continuum limit
whereas non-Abelian gauge theories are chaotic in the continuum,
although the exact scaling relation is still an open problem.
Chaos to order transitions were observed in a spatially
homogeneous SU(2) Yang-Mills-Higgs system and in a spatially homogeneous
SU(2) Yang-Mills Chern-Simons Higgs system \cite{Sala97,Mukk97}.
In Ref.~\cite{Sala97} a chaos to order transition was also seen
on the quantum level, i.e. a smooth transition from a Wigner
to a Poisson distribution was found. 
A transition in $P(s)$ from Wigner to Poisson behavior was further
observed at the metal-insulator transition of the Anderson model
\cite{Alts88}.
Further, the suppression of the characteristic manifestations
of dynamical chaos by quantum fluctuations was analyzed in 
the context of spatially homogeneous scalar electrodynamics \cite{Mati97}
and for a $0+1$-dimensional space-time $N$-component $\phi^4$ theory
in the presence of an external field \cite{Case98}.
These chaos to order transitions were seen in spatially
homogeneous models and not for the full classical field theory.
The relationship to properties of the quantum field theory is an
interesting issue.

Here we focus on the Dirac operator for quenched $4d$ compact 
quantum electrodynamics (QED)
to search for the possible existence of a transition from chaotic to
regular behavior in Abelian lattice gauge
theories. In particular, we are interested in the 
nearest-neighbor spacing distribution of the eigenvalues of the Dirac 
operator across the phase transition from the strong coupling to the 
Coulomb phase. In the strong coupling region Abelian as well as non-Abelian 
lattice gauge theories are in a confined phase \cite{Wils74}. 
For compact QED this means that for couplings 
$\beta < \beta_c \approx 1.01$ the electron is confined.
However, when crossing the phase transition the conventional Coulomb phase is
observed.  It is an interesting question if the difference between
the Coulomb phase in QED and the quark-gluon plasma phase in QCD has an
influence on the level repulsion of the corresponding Dirac spectra.

\subsection{Quantum chaos of fermion fields}

We generated gauge field configurations using the standard Wilson plaquette
action for U(1) gauge theory,
\begin{equation}
S_G(U_l)=\beta\sum_P\nolimits \, (1-\cos \Theta_P) \; ,
\end{equation}
where $U_l\equiv U_{x,\mu}=\exp(i\Theta_{x,\mu})$, with $\Theta_{x,\mu}\in
[-\pi,\pi)$, are the field variables defined on the links $l\equiv(x,\mu)$.
The plaquette angles are $\Theta_P = \Theta_{x,\mu}+\Theta_{x+\hat{\mu},\nu}
-\Theta_{x+\hat{\nu},\mu}-\Theta_{x,\nu}$. We simulated $8^3 \times 4$
and $8^3 \times 6$ lattices at various values of the inverse gauge
coupling $\beta=1/g^2$
both in the strong coupling and the Coulomb phase. Typically we discarded
the first 10000 sweeps for reaching equilibrium and produced 20 
independent configurations separated by 1000 sweeps for 
each $\beta$. Because of the spectral ergodicity property of RMT one can
replace ensemble averages by spectral averages \cite{Guhr98} if one is
only interested in the bulk properties. Thus a few independent
configurations are sufficient to compute $P(s)$.

On the Euclidean lattice the Dirac operator $/\!\!\!\!D=/\!\!\!\partial+ig\: /\!\!\!\!A$ 
for staggered fermions 
\begin{equation}
M_{x,x'}=\frac{1}{2}\sum\limits_{\mu=1}^{4}\eta_{x\mu} \left( 
\delta_{x+\hat{\mu},x'} U_{x,\mu} - \delta_{x-\hat{\mu},x'} 
U_{x,\mu}^{\dagger} \right) 
\end{equation}
is anti-Hermitian so that all 
eigenvalues are imaginary. For convenience we denote them by 
$i\lambda_n$ and refer to the $\lambda_n$ as the eigenvalues in the 
following. Because of $\{/\!\!\!\!D,\gamma_5\}=0$ the $\lambda_n$ 
occur in pairs of opposite sign.  All spectra were checked against 
the analytical sum rules
\begin{equation}
\sum_{n} \lambda_n = 0 \qquad {\rm and} \qquad
\sum_{\lambda_n>0} \lambda_n^2 = V \:,
\end{equation}
where $V$ is the lattice volume.

\begin{figure}[t]
\centerline{{\psfig{figure=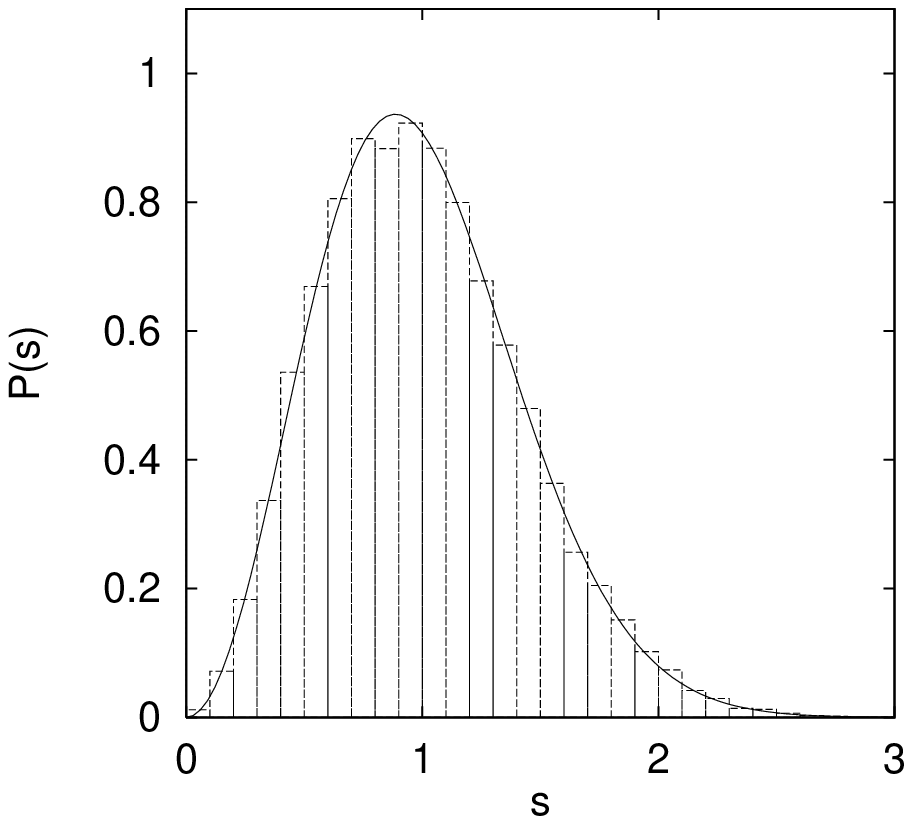,width=6cm}}\hspace{5mm}
{\psfig{figure=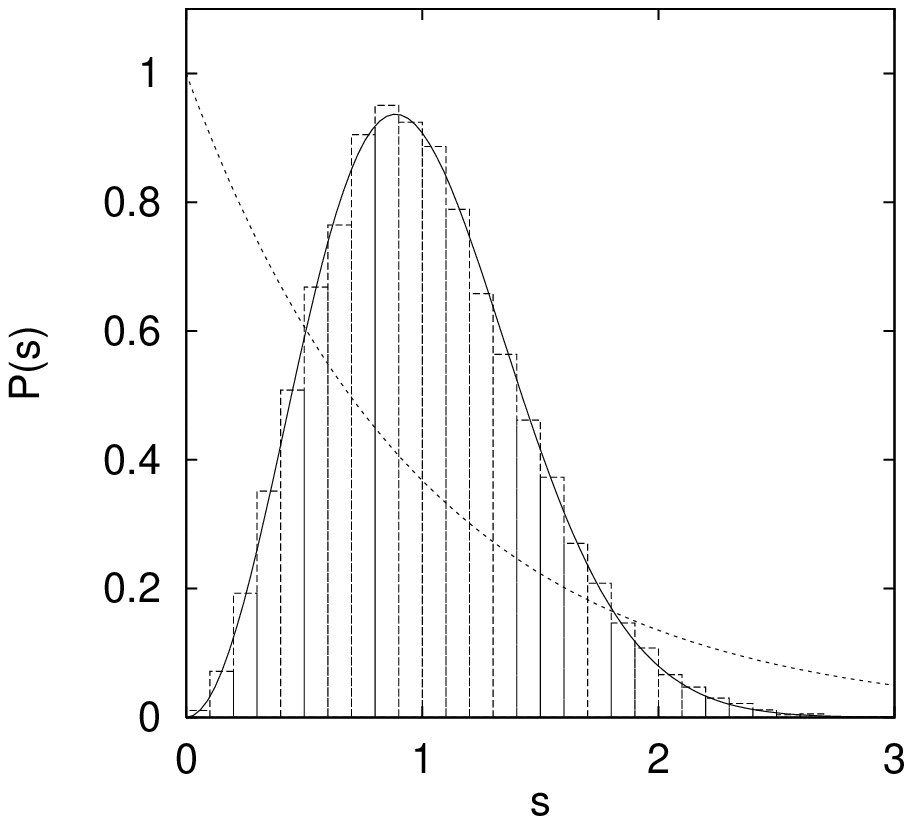,width=6cm}}}
  \caption{\label{fig4} Nearest-neighbor spacing distribution $P(s)$
       of the Dirac operator for compact U(1) theory in the strong
       coupling phase for $\beta=0.90$ (left) and in the Coulomb
       phase for $\beta=1.10$ (right).
       The histogram represents the lattice data on an $8^3 \times 6$
       lattice averaged over 20 independent configurations. The full
       curve is the Wigner distribution of
       Eq.~(\protect\ref{ue}) for the unitary ensemble of RMT.
       For comparison the Poisson distribution
       $P(s) = e^{-s}$ is also indicated by the dashed line.}
\end{figure}

\begin{figure}[t]
  \centerline{\hbox{
  \psfig{figure=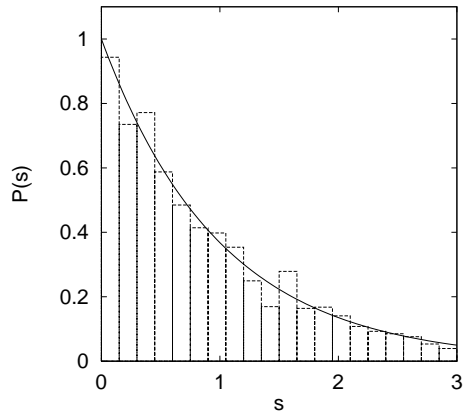,width=6cm}
  }}
  \caption{\label{fig5} Nearest-neighbor spacing distribution $P(s)$
       of the analytically calculated eigenvalues of
       Eq.~(\protect\ref{free}) for a free Dirac operator on a
       $53\times 47 \times 43 \times 41$ lattice (histogram)
       compared with the Poisson distribution $P(s) = e^{-s}$
       (solid line).}
\end{figure}

In RMT one has to distinguish between different universality classes
which are determined by the symmetries of the system. So far the 
classification for the QED Dirac operator has not been done. Our
calculations show that in the case of the staggered $4d$ compact QED Dirac matrix
the appropriate ensemble is the unitary ensemble. Although from a 
mathematical point of view this is the simplest one, the RMT result
for the nearest-neighbor spacing distribution is still rather complicated.
It can be expressed in terms of so-called prolate spheroidal functions,
see Ref.~\cite{Meht91} where $P(s)$ has also been tabulated. A good
approximation to $P(s)$ is provided by the Wigner surmise for the 
unitary ensemble
\begin{equation}
  \label{ue}
  P(s)=\frac{32}{\pi^2}\,s^2\,e^{-\frac{4}{\pi}s^2}\:.
\end{equation}

We have simulated $8^3 \times 4$ lattices at $\beta=0,\:0.90,\:0.95,\:1.00,\:
1.05,\:1.10,\:1.50$ and $8^3 \times 6$ lattices at $\beta=0.90,\:1.10,\:
1.50$. All results are similar to those selected for the plots.
The left plot in Fig.~\ref{fig4} shows the 
nearest-neighbor spacing distribution $P(s)$ for $\beta=0.90$ in the
confined phase averaged over 20 independent configurations on the
$8^3 \times 6$ lattice compared with the Wigner surmise for the unitary
ensemble of RMT of Eq.~(\ref{ue}). Good agreement is found. According
to the Bohigas-Giannoni-Schmit conjecture this means the system can be
regarded as chaotic in the strong coupling region. The right plot
in Fig.~\ref{fig4} shows the nearest-neighbor spacing distribution 
$P(s)$ for $\beta=1.10$ in the Coulomb phase again averaged over 
20 independent configurations and compared
with the Wigner surmise (\ref{ue}).
The agreement of the lattice data with the RMT predictions is
interpreted as a signal that quantum chaos survives the phase transition.
We find no deviation up to the maximum coupling considered, $\beta = 1.50$. 

In the strong coupling phase the result holds down to $\beta=0$.
Therefore, we tend to interpret our, as well as previous 
\cite{Pull98,Verb95}, results in the sense that the disorder of 
the gauge field configurations \cite{Biro94,SU2} is responsible
for the chaotic characteristics of the spectrum of the Dirac operator.
In contrast 
to that: The free fermion theory is non-chaotic and the
corresponding
nearest-neighbor spacing distribution obeys a Poisson distribution.
This is illustrated in Fig.~\ref{fig5} where $P(s)$ is 
obtained from the analytical eigenvalues of the free 
Dirac operator on a $53\times 47\times 43 \times 41$ lattice:
\begin{equation}
\label{free}
a^2\lambda^2 = \sum\limits_{\mu=1}^4 \sin^2 \left( \frac{2\pi n_{\mu}}{L_{\mu}}
\right)\:.
\end{equation}
Here $a$ is the lattice constant, $L_{\mu}$ is the number of lattice sites
in $\mu$-direction, and $n_{\mu}=0,...,L_{\mu}-1$. We used an asymmetric 
lattice with $L_{\mu}$ being primes and restricted the range to
$(L_{\mu}-1)/2$ instead of $L_{\mu}-1$ in each direction to avoid 
degeneracies of the free spectrum.

\section{Conclusion}

In the underlying article we performed a comparison of classical
chaos and quantum chaos in fundamental field theories of physics,
exemplified for the U(1) theory of electrodynamics. This is not a 
direct comparison, however, since it deals with the gauge field
in classical theory and with fermions in the quantum case. It
turned out that the classical U(1) field is chaotic in the confinement
phase with decreasing Lyapunov exponents towards the Coulomb phase.
A scaling analysis indicates a regular continuum theory as one expects 
from the Maxwell equations. On the other hand, our investigation of
the quantized fermion field fulfills the criterion for quantum
chaos both in the confinement and Coulomb phase. A scaling analysis
was not possible for the quantum case (due to the lack of a 
$\beta$-function) which could cover the transition to a regular
theory. Nevertheless, the free Dirac operator, in absence
of a covariant derivative and minimal gauge coupling, exhibits
regular behavior.

It would be interesting to study the direct counterpart of 
the classical gauge field after quantization.
A similarly accurate determination
of the eigenvalue spectrum of the gauge sector necessitates
to construct the corresponding Fock space and to diagonalize
high-dimensional matrices which seems to be out of reach
for $4 d$ QED/QCD. On the other hand, chaos studies of the
classical limit of the fermion field would also be of 
interest but have not yet been attempted.

{\bf Acknowledgments.  }
This work has been supported by the Austrian National Scientific Fund under
the project FWF P14435-TPH.
We thank Bernd A. Berg, Tam\'as S. Bir\'o, Natascha H\"ormann 
and Wolfgang Sakuler for previous cooperations.

\medskip

\end{document}